\documentclass[prl,showpacs,amsfonts,amsmath,twocolumn]{revtex4}
\usepackage[sort&compress]{natbib}

\usepackage[final]{graphicx}
\newlength\figurewidth
\AtBeginDocument{\setlength\figurewidth{.9\linewidth}}

\def\kB{k_{\text{B}}}
\def\vec#1{{\boldsymbol #1}}

\begin{document}
\title{Idealized glass transitions under pressure:
dynamics versus thermodynamics}
\date{\today}
\def\edin{\affiliation{University of Edinburgh, School of Physics,
  Mayfield Road, Edinburgh EH9 3JZ, U.K.}}
\def\dlr{\affiliation{Institut f\"ur Materialphysik im Weltraum,
  Deutsches Zentrum f\"ur Luft- und Raumfahrt (DLR), 51170 K\"oln,
  Germany}}
\author{Th.~Voigtmann}\dlr

\begin{abstract}
The interplay of slow dynamics and thermodynamic features of
dense liquids is studied by examinining how the glass transition
changes depending on
the presence or absence of Lennard-Jones-like attractions.
Quite different thermodynamic behavior leaves the dynamics unchanged,
with important
consequences for high-pressure experiments on glassy liquids.
Numerical results are obtained within
mode-coupling theory (MCT), but the qualitative features are argued to hold
more generally.
A simple square-well model can be used to explain generic features
found in experiment.
\end{abstract}

\maketitle


The quest for identifying the physical mechanism behind the dynamical
transformation of a liquid into an amorphous solid, the glass
transition, has prompted many studies aiming to disentangle the dominant
control variables involved. It is recognized that the slow dynamics
connected with the glass transition is universal, but
its connection to the underlying liquid structure is highly debated
\cite{Ito.1999,Sastry.2001,Bordat.2004,Novikov.2004,Yannopoulos.2006}.
Some argue in terms of a density effect called free or excluded
volume; others attribute the main physics to energetic interactions
and thermally activated processes.
%

Experiments changing both temperature $T$
and pressure $P$ close to the transition
are emerging to resolve such issues, but
have brought contradictory results.
Some find that temperature dominates glassy dynamics by far
\cite{Ferrer.1998,Hensel.2002b,Dreyfus.2003,Tarjus.2004};
some that it does not 
\cite{Paluch.2002b,Paluch.2002c,Barbieri.2004}.
Others find both variables to exert
equal influence
\cite{
Corezzi.1999,Comez.2002,Casalini.2002,Paluch.2000,Li1995,%
Paluch.2002f,Paluch.2003,Roland.2003,Mpoukouvalas.2003,Paluch.2003c},
some with temperature
\cite{Leyser.1995,
Roland.2003c,Casalini.2003f,Casalini.2003d,%
Papadopoulos.2004},
some with density $\varrho$ being more relevant
\cite{Cook.1994,Paluch.2003b,Barbieri.2004b,Roland.2004}.
The difficulty of obtaining data over wide pressure
ranges might be impeding:
few studies, pioneered only in the 1990's
\cite{Cook.1993,Cook.1994}, go beyond $1\,\text{GPa}$.

Here we propose that to resolve the apparent contradictions, one needs
to \emph{separate}
non-universal thermodynamic aspects, namely the equation of state (EOS) of
the system, from universal dynamical features, viz.\ the slow relaxation.
Specifically, we show how the presence or absence of attractive
interactions affects the glass transition,
and how this emerges in different pairs
of variables linked by the EOS: $(\varrho,T)$ (preferred by
theory) vs.\ $(P,T)$ (more amenable to experiments), yielding a
transition that appears `temperature-driven' in the latter.


The Lennard-Jones (LJ) potential serves as a realistic
interaction model:
$V_{\text{LJ}}(r)=4\epsilon[(r/\sigma)^{-12}-(r/\sigma)^{-6}]$,
with dimensionless parameters
$\varrho^*=\varrho\sigma^3$, $T^*=1/(\beta\epsilon)$,
$P^*=P\sigma^3/\epsilon$;
$\beta=1/(\kB T)$ with Boltzmann's constant $\kB$.
To study relative effects of entropy and energy,
we compare the LJ glass transition
with that of its purely repulsive (LJR) counterpart,
$V_{\text{LJR}}(r)=V_{\text{LJ}}(r)+\epsilon\ge0$ for $r\le2^{1/6}\sigma$,
$V_{\text{LJR}}(r)=0$ else.
For the transition lines, mode-coupling theory (MCT)
\cite{Bengtzelius1984,Goetze1991b} provides a reasonable qualitative
description. Its transition temperature
$T_c$ is systematically above the
experimental (calorimetric) one, $T_g$,
but nevertheless serves as a good indicator for the change from
high-$T$ liquid-like dynamics to low-$T$ glass-like one
\cite{Novikov2003,Meyer2002}. In MCT,
%
$T_c$ is the point where the glass form factor $f(q)$, measuring an
elastic response to the scattering spectrum, jumps discontinuously (usually
from zero to a finite value) due to a bifurcation in
\cite{Goetze1991b}
\begin{equation}\label{mctintegral}
 \frac{f(q)}{1-f(q)}=\frac{\varrho S(q)}{2q^4}
  \int\frac{d^3k}{(2\pi)^3}
  {\mathcal V}(\vec q,\vec k) f(k)f(p)\,,
\end{equation}
${\mathcal V}(\vec q,\vec k)=S(k)S(p)
(q^2-\vec q\vec k/S(k)-\vec q\vec p/S(p))^2/\varrho^2$,
with wave number $q=|\vec q|$, $p=|\vec q-\vec k|$.
Equation~\eqref{mctintegral} is solved numerically by an iteration scheme
\cite{numericsnote}.
The interaction potential $V(r)$ and temperature $T$ enter only through
the static structure factor $S(q)$, obtained
in the hypernetted-chain (HNC)
approximation \cite{Hansen1986,Labik}.
%
$S(q=0)$ determines the pressure through the EOS,
\begin{equation}\label{eos}
  \beta P=\int_0^\varrho d\hat\varrho\,S(q=0,\hat\varrho,T)^{-1}\,,
\end{equation}
integrated numerically
for the LJ and LJR systems.
We have checked that the quantitative error of the HNC closure does
not influence our results qualitatively.


\begin{figure}
\includegraphics[width=\figurewidth]{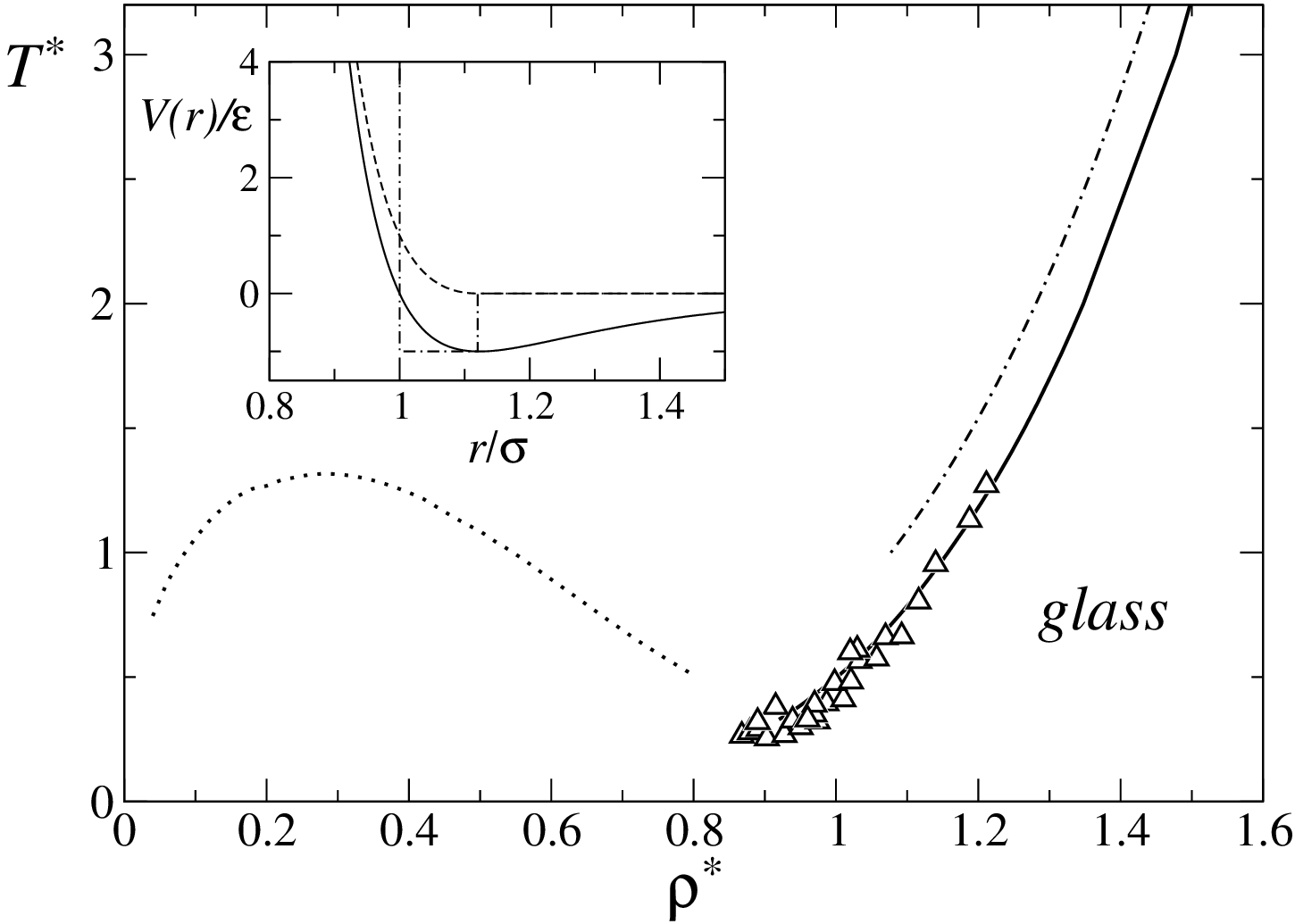}
\caption{\label{rhoTplot}
  Glass transition lines for the Lennard-Jones system with and without
  attractions: MCT results for both coincide
  (solid line). Triangles are glass-transition points from
  simulations \protect\cite{Bengtzelius1986b,Leonardo.2000}.
  Dash-dotted line: soft-sphere asymptote,
  $T^*_c(\varrho)\approx(\pi/6\varphi_{\text{eff}}^c)^4\varrho^4$,
  $\varphi_{\text{eff}}^c=0.564$.
  The dotted line indicates a gas-liquid spinodal.
  Inset: LJ (solid line), LJR (dashed), and
  square-well ($\delta=0.12$; dash-dotted) potentials.
}
\end{figure}

The MCT glass transitions for the two systems
are almost identical, i.e., they depend very little on the
presence or absence of attractive interactions. As shown in
Fig.~\ref{rhoTplot}, the lines in a $T$-$\varrho$ diagram coincide on
the scale of the figure. In fact, most of the change in $T_c(\varrho)$
can be understood from the soft-sphere limit:
for $T\to\infty$, $V_{\text{LJ(R)}}(r)\sim\epsilon(r/\sigma)^{-n}$, $n=12$,
and the only control parameter is an effective packing fraction,
$\varphi_{\text{eff}}=(\pi/6)\varrho\sigma_{\text{eff}}^3$, where
$\sigma_{\text{eff}}^3=\sigma^3\left.T^*\right.^{-3/n}$ accounts for the
soft core.
The dash-dotted line in Fig.~\ref{rhoTplot} corresponds to this soft-sphere
glass transition, $\varrho_c\propto T_c^{-3/n}$,
where Eq.~\eqref{mctintegral} yields
$\varphi_{\text{eff}}^c\approx0.564$. It clearly shows the same qualitative
behavior as the $T_c(\varrho)$ lines for the full LJ system.
Molecular-dynamics simulation data on the
LJ glass transition collected in Ref.~\cite{Bengtzelius1986b} and estimates
from energy-landscape calculations of
Ref.~\cite{Leonardo.2000} are reproduced in
Fig.~\ref{rhoTplot} as triangles. They agree with our results reasonably
well, considering the different definitions of the glass-transition point
used in these works.

Attractions have a crucial effect on the thermodynamics: they introduce
a gas-liquid spinodal, whose estimate
is shown in Fig.~\ref{rhoTplot} (dotted line).
The compressibility diverges smoothly there, $1/S(q=0)\to0$.
Hence, the resulting $P(\varrho,T)$ values in the liquid,
Eq.~\eqref{eos}, are significantly
lower with than without attractions,
i.e., for $T<T_{\text{cr}}$ and $\varrho>\varrho_{\text{cr}}$,
where $(\varrho_{\text{cr}},T_{\text{cr}})$ is the gas-liquid critical point.
This even holds for approximations that fail to
predict the spinodal (such as HNC), but replace it with branch points
where $S(q=0)$ is large
\cite{Belloni}.
In other words, LJ-like attractions in
high-density liquids simply affect the pressure:
the contributions of all particles add up to a
flat background that does not influence the forces \cite{Widom1967},
nor the (glassy) dynamics.

\begin{figure}
\includegraphics[width=\figurewidth]{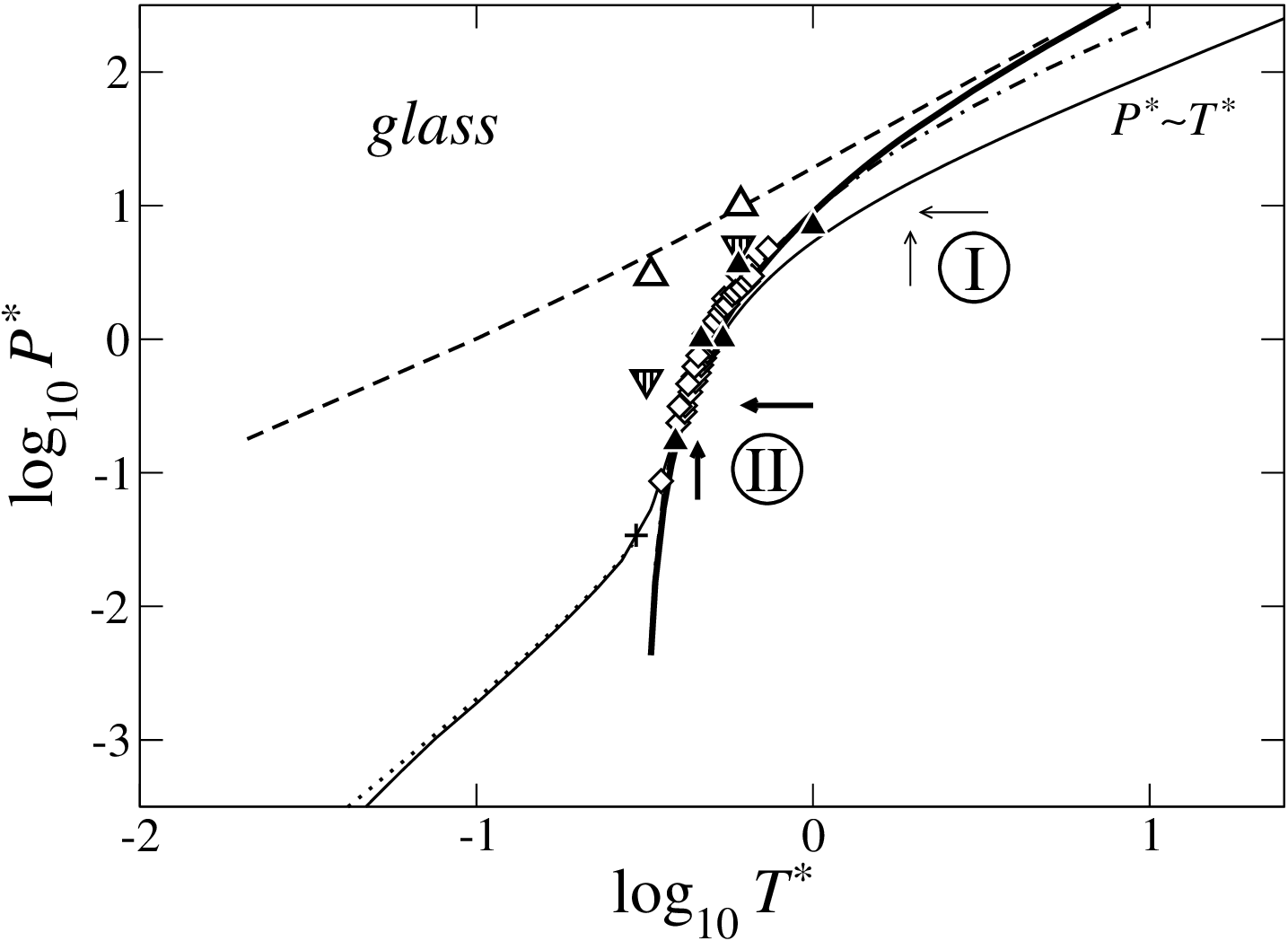}
\caption{\label{PTplot}
  Lennard-Jones
  glass transitions with (thick line) and without (dashed) attractions in
  a pressure-vs-temperature representation, from MCT.
  Triangles: simulation data
  from Refs.~\protect\cite{Bengtzelius1986b,MaLai} (filled: LJ; open: purely
  repulsive; inverted: truncated LJ). Dash-dotted line: LJ transition
  as $P^*_{\text{eff}}=P^*T^{-1/4}$.
  Thin solid line: square-well model transition
  ($\delta=0.12$, shifted as $T^*\mapsto 1.5T^*$);
  dotted: square-well model spinodal with critical point (cross).
  Diamonds:
  experimental data for glycerol (Refs.~\protect\cite{Cook.1994,Paluch.2002}),
  see text. Arrows indicate isobars and
  isotherms used in Fig.~\protect\ref{viscplot}.
}
\end{figure}

Figure~\ref{PTplot} demonstrates the marked effect the spinodal has on the
glass transition
line in the $P$-$T$ diagram: although nearly indistinguishable in the
$T$-$\varrho$ diagram, Fig.~\ref{rhoTplot}, the transition lines
with (LJ) and without (LJR) attraction
(thick solid and dashed lines) now appear qualitatively different.
The LJ $\log P_c$-versus-$\log T$ line has a steep part not present in
the LJR line at $\log_{10}T^*<0.1$, corresponding to
$T^*<T^*_{\text{cr}}\approx1.4$.
It stops at the spinodal,
restricting the glassy regime to densities $\varrho>\varrho_{\text{cr}}$
\cite{Ashwin.2006}.
Simulation data
from Refs.~\cite{Bengtzelius1986b,MaLai} (triangles in Fig.~\ref{PTplot})
scatter between the LJ and LJR lines since the simulations used different
truncations of the potential, yielding different equations of state.
The data from Ref.~\cite{Bengtzelius1986b}, where full
$(\varrho,T,P)$ triplets are given for several truncations, collapse
to a single curve in Fig.~\ref{rhoTplot} within error bars,
confirming our finding.

The above results are easily
understood within MCT: 
the glass transition
in dense systems is driven by nearest-neighbor interactions (the
cage effect), i.e., by features of the structure factor $S(q)$ at
$q\approx2\pi/\sigma$.
But the presence of attractions affects only $S(q\to0)$ \cite{WCA}; a region
strongly suppressed in the MCT integral,
Eq.~\eqref{mctintegral}. Merely the transformation $\varrho\mapsto P$ is
dominated by $q\to0$ effects.

Since this transformation in general relies on straightforward but cumbersome
numerical calculations, we simplify matters by introducing a square-well (SW)
potential as a `cartoon' of the LJ model, showing first that the qualitative
features discussed above are still preserved. The SW model consists
of a hard-sphere core and an attraction of relative width
$\delta$:
$V_{\text{SW}}(r)=-\epsilon$ for $1<r/\sigma<1+\delta$.
Here, a
mean-spherical approximation (MSA) for $S(q)$ \cite{Dawson2001} allows
to integrate Eq.~\eqref{eos} analytically, greatly simplifying
calculations. The MCT line
for $\delta=0.12$ \cite{numericsnote}
is shown in Fig.~\ref{PTplot},
rescaled as $T^*\mapsto 1.5T^*$ to account for the different $S(q)$
approximation that mainly induces a shift in the $T$ scale \cite{Dawson2001}.

We identify two generic regimes for
both the LJ and the square-well lines:
$T^*\to\infty$ is the hard-sphere
limit where the glass transition occurs along an
isochore $P_c\sim\kB T$;
this limit is approached for $T^*\gtrsim1$ (regime~I). It
is also present in the LJ system, provided one corrects for
soft-core effects, $P_{\text{eff}}^*
=P^*T^{-3/n}$, as the dash-dotted line shows.
For $0.1\lesssim T^*\lesssim1$ the steep
$\log P_c^*$-versus-$\log T^*$ line discussed above is found
in both models (regime~II).
Its position along the $T^*$ axis
scales with the gas-liquid critical temperature
($T^*_{\text{cr}}\sim\delta$ for the MSA-SW).
For $T^*\to0$, the SW model shows a low-density regime
corresponding to densities
$\varrho<\varrho_{\text{cr}}$ (a cross marks the point where $\varrho_c
=\varrho_{\text{cr}}$), where
the dynamics itself strongly depends on the
potential depth $\epsilon$. This attraction-driven regime may be
connected with colloidal gelation and is absent in the
Lennard-Jones model and in common molecular glasses.
Relevant for typical glass formers
at $\text{MPa}$ pressures is regime~II, as
pointed out in a recent study
\cite{paperI}: experiments reveal steep $P(T_g)$ curves that are
incompatible with the hard-sphere-like regime~I. Diamond symbols in
Fig.~\ref{PTplot} exemplify
this for glycerol:
experimental $T_g(P)$ data from Refs.~\cite{Cook.1994,Paluch.2002}
was mapped according to $\epsilon/\kB=500\,\text{K}$ and
$\epsilon/\sigma^3=2.5\,\text{GPa}$, just to demonstrate qualitative
agreement (and absorbing the quantitative difference between $T_g$ and $T_c$
in the mapping).



Our discussion of glass-transition lines has direct
implications for dynamical quantities, as the latter are expected to depend
strongly on the distance to the transition. For example, a `thermodynamic
scaling' has been observed for many glass formers, that
involves $\varrho T^{-\gamma}$ as a scaling variable
\cite{Dreyfus.2003,Casalini.2004,Roland.2004,%
Pawlus.2004,Roland.2005b} and can be interpreted as an effective
density-dependent interaction energy \cite{Tarjus.2004}.
$\gamma$ is an empirical, effective exponent: even in the LJ system,
$\gamma\neq1/4=3/n$ \cite{Coslovich.2008}. In agreement with this, we find
that the LJ transition line can be well fitted for all $T^*<3$ by
$\varrho^*_c(T)\propto T^\gamma$, where $\gamma=0.15\ldots0.23$.
Partly, this merely mirrors that effective power laws can be used to fit
the potential in the respective $V(r)/\epsilon\approx T^*$ range.
If attractions are present, the effective $\gamma$ also depends on their
details, as strikingly demonstrated by the SW system, where we do not
find the $\gamma\to0$ expected from the $n\to\infty$ hard-sphere limit.

\begin{figure}
\includegraphics[width=\figurewidth]{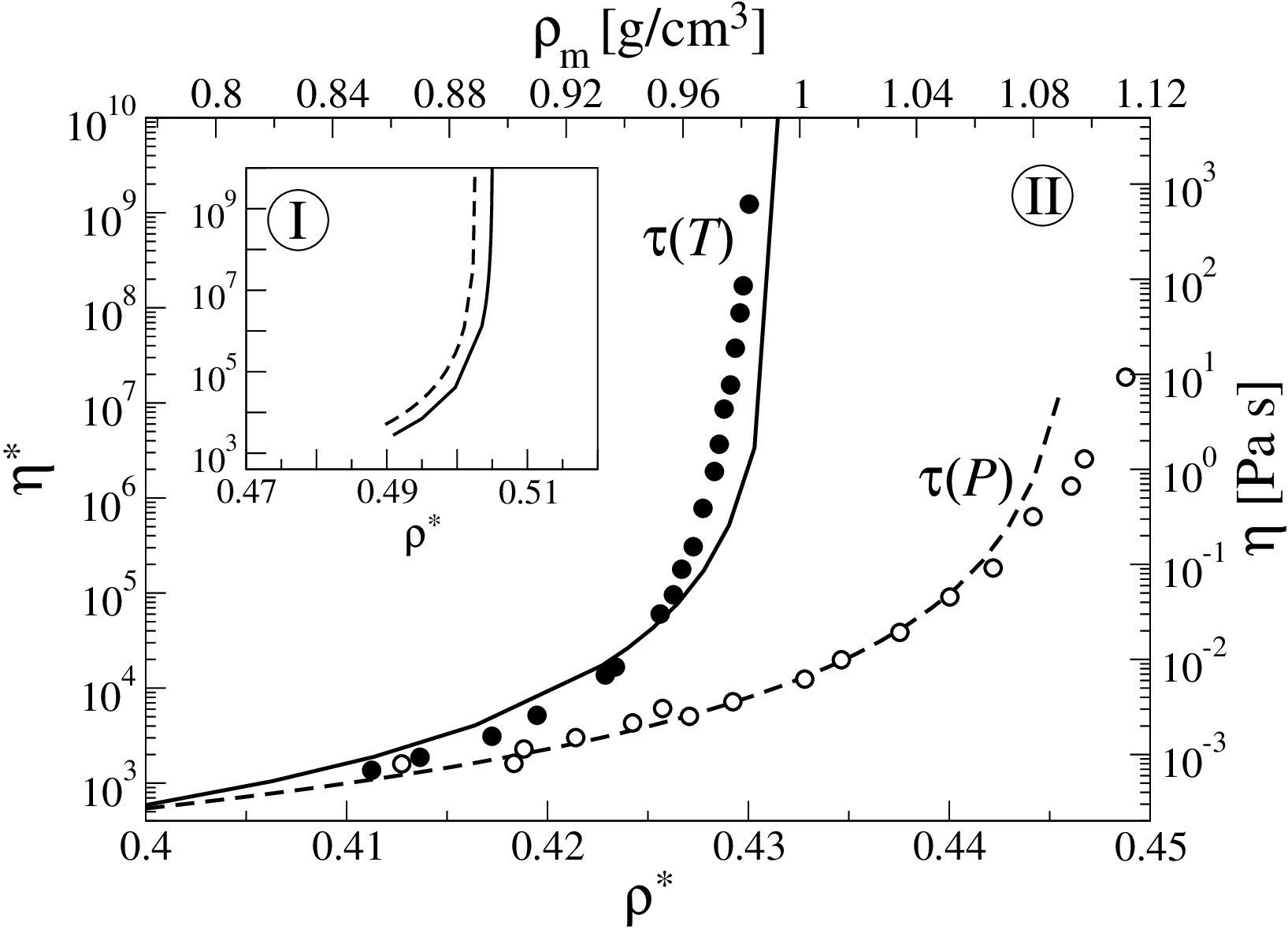}
\caption{\label{viscplot}
  Typical relaxation times $\tau$ as functions of density:
  MCT results for the SW system ($\delta=0.12$; bottom and left
  axes), along the isobar $P^*=0.316$ (solid line) and the isotherm
  $T^*=0.3$
  (dashed), marked by thick arrows in Fig.~\protect\ref{PTplot}.
  Symbols (upper/right axes) are viscosity data for isopropylbenzene
  \protect\cite{Li1995} for the isobar $P=0.1\,\text{MPa}$ (filled)
  and the isotherm $T=293\,\text{K}$ (open).
  Inset: MCT results for $P^*=8.91$ (solid)
  and $T^*=0.775$ (dashed); thin arrows in Fig.~\protect\ref{PTplot}.
}
\end{figure}

One can characterize the relative effects of temperature and pressure on the
glass transition
by monitoring the viscosity $\eta$ 
along isotherms and isobars, using
density as a parameter. Experiments
\cite{Cook.1994,Leyser.1995,Li1995,Ferrer.1998,Paluch.2002b,Paluch.2002c,%
Paluch.2002f,Paluch.2003,Paluch.2003b,Mpoukouvalas.2003,Paluch.2003c,%
Roland.2003,Roland.2003c,
Casalini.2003d,%
Casalini.2003f,Papadopoulos.2004}
usually find stronger variation along an isobar (varying $T$)
than along an isotherm (varying $P$).
This is a natural consequence of our scenario,
evidenced in Fig.~\ref{viscplot}
by the viscosity $\eta^*$ in SW units, calculated from the standard
Green-Kubo expression in MCT approximation \cite{Bengtzelius1984}.
Consider regime~II:
changing $T$ along the $P^*=0.316$ isobar (solid line)
corresponds to a more direct approach to the glass transition line
as compared to changing
$P$ along the $T^*=0.3$ isotherm (dashed);
cf.\ the thick arrows in Fig.~\ref{PTplot}.
Clearly, $\eta^*$
diverges over a smaller density interval along the isobar.
The agreement with experiment is semi-quantitative, as the symbols,
reproduced from Ref.~\cite{Li1995}, demonstrate.
Only in regime~I (inset of Fig.~\ref{viscplot};
thin arrows in Fig.~\ref{PTplot}) do $P$ and $T$
excert equal influence.
Not all transition
points are equal: MCT predicts nonuniversal
amplitudes and shapes for relaxation spectra that change along
the transition line, invalidating a `temperature--pressure'
superposition principle. But the changes are small enough to make it
appear to work, explaining why some experiments find it
\cite{
Corezzi.1999,Paluch.2002,
Comez.2002,Casalini.2002,Roland.2003c,
Barbieri.2004},
some with restrictions
\cite{
Hensel.2004,Sekula.2004,Papadopoulos.2004,Niss.2007},
some not at all \cite{Patkowski.2003,Casalini.2003f}.

Temperature- and density-effects are often quantified
by a ratio of activation energy and enthalpy \cite{Williams.1964,
Ferrer.1998,Roland.2004},
$E_V/H_P=(\partial\log\tau/\partial T^{-1})_V/
(\partial\log\tau/\partial T^{-1})_P$, trivially related to
$E=(\partial\varrho/\partial T)_\tau(\partial T/\partial\varrho)_P$.
Large $|E|$ signify a temperature-driven transition, small $|E|$ a
density-driven one. $E$
consists of a glass-transition part, and a purely EOS-driven one.
According to Fig.~\ref{rhoTplot}, the isokinetic term
$r=(\partial\varrho/\partial T)_\tau$ does not depend on LJ-like attraction;
the EOS term $t=(\partial T/\partial\varrho)_P$ however changes.
Estimating the latter through the SW-MSA expression, we find that in fact,
$t$ \emph{decreases} in the vicinity of the spinodal: the measure
$|E|$ increases with $P$, indicating a growing influence of temperature at
higher pressures.
Such trends have been found in experiments and argued to be at odds
with the expectation that the transition becomes hard-sphere like at high $P$
\cite{Tarjus.2004}. According to our model, they are dominantly
thermodynamic effects.

A similar conclusion holds for the pressure dependence of `fragility'
often used to classify how quickly relaxation times diverge.
Recent work debated its relation to $q\to0$ quantities
such as elastic constants, the above energy-enthalpy ratio or
the effective exponent $\gamma$
\cite{Ito.1999,Sastry.2001,Bordat.2004,Novikov.2004,Yannopoulos.2006,%
Niss.2007,Roland.2008}.
If true, MCT predicts a pressure dependent fragility that is a nonuniversal
feature of $S(q\to0)$, less so of the glassy dynamics \cite{fragility}.
This may also hint towards why the presence of hydrogen bonds drastically
changes the high-pressure behaviour regarding fragility and
thermodynamic scaling \cite{Roland.2008}.


In conclusion, MCT glass-transition lines for the Lennard-Jones system and
for the same system truncated to be purely repulsive are nearly
identical in a $\varrho$-$T$ plot, Fig.~\ref{rhoTplot}. Yet,
they appear quite different in a $P$-$T$ diagram,
Fig.~\ref{PTplot}. The difference can be understood as unrelated to
the glass-transition mechanism itself, but to a
difference in thermodynamic behavior only.
If one accepts that the glass transition is a primarily dynamic phenomenon,
it will not be altered by sufficiently long-ranged, LJ-like, attractions.
However, the equation of state, determining the pressure of the system,
will change.
The $P_c$-versus-$T$ curve hence has two regimes if attractions are present.
In the very-high pressure regime I, it is essentially
a density-driven fluid-glass transition:
$P_c\propto T$, after correcting for soft-core effects.
In regime II, identified as the experimentally relevant one,
the existence of a gas-liquid spinodal leads to $T_c(P)$ curves with
a much weaker $P$-dependence:
this could be called a ``temperature-driven'' liquid-glass
transition; but ``temperature driven'' is not equivalent to
``attraction dominated''.
Key qualitative features can be understood with the help of the
square-well system as a better tractable model.

Discussing the glass transition in terms of
``temperature vs.\ pressure'' might obscure the physics
responsible for it, focusing too much on
different thermodynamics of different glass formers.
For example, a change in composition in metallic glass formers
greatly changes the thermodynamics, but has little
effect on the slow dynamics \cite{MavilaChathoth2004}.
Experiments probing the pressure range $P\gg1\,\text{GPa}$
seem desirable to test the picture proposed here.

\begin{acknowledgments}
I thank M.~Sperl, M.~E.~Cates, W.~C.~K.~Poon, A.~Meyer, and W.~G\"otze
for discussions
and EPSRC (GR/S10377) and DFG (Vo-1270/1) for funding.
\end{acknowledgments}

\bibliography{lit}

\begin{thebibliography}{61}
\bibitem[{num()}]{numericsnote}
\bibinfo{note}{$M=200$ $q$-points with cutoff $Q\sigma=80$ are used; for the
  square-well model, $M=600$ and $Q\sigma=240$.}
\end{thebibliography}
\bibliographystyle{apsrev-etal}

\end{document}